# Ion-Hose Instability in Long Pulse Induction Accelerators

George J. Caporaso and Jim. F. McCarrick
Lawrence Livermore National Laboratory, Livermore, California 94550 USA

*Abstract*

The ion-hose (or fast-ion) instability sets limits on the allowable vacuum in a long-pulse, high current accelerator. Beam-induced ionization of the background gas leads to the formation of an ion channel which couples to the transverse motion of the beam. The instability is studied analytically and numerically for several ion frequency distributions. The effects of beam envelope oscillations on the growth of the instability will be discussed. The saturated non-linear growth of the instability is derived analytically and numerically for two different ion frequency distributions.

## 1 INTRODUCTION

With the advent of DARHT-2 and its 2 μsec pulse concern has surfaced over the ion-hose instability that may arise from the beam's interaction with the channel created from the background gas via collisional ionization [1]. We consider a simple model with a τ-dependent neutralization (τ here is defined as the distance back from the beam head divided by βc). The model we use for the channel is really more appropriate for a channel that has been preformed however a more correct treatment (which is also more complicated) produces the same asymptotic result so we adopt this one for simplicity. Let y represent the centroid position of the beam and ψ represents the centroid position of the ion channel. We consider the case of a smooth external focusing force (the case of solenoidal focusing is well represented in the asymptotic solutions by replacing $k_\beta$ the betatron wavenumber by $k_c/2$, one half of the cyclotron wavenumber).

Since we will also follow the non-linear development of the instability we choose as a starting point the equations used by Buchanan to describe the coupling of a beam and channel, each assumed to have a Gaussian spatial profile [2]. The beam has Gaussian radius a while the channel has Gaussian radius b such that the parameter $R_o$ is given by

$$R_o^2 \equiv a^2 + b^2. \quad [1]$$

This parameter results from integrating the force due to the beam over the distribution of the channel and vice versa. The model is

$$\frac{\partial^2 \hat{y}}{\partial \varsigma^2} + \hat{y} + \varepsilon x \alpha^2 (\hat{y} - \hat{\psi}) = 0 \quad [2]$$

$$\frac{\partial^2 \hat{\psi}}{\partial x^2} + \alpha^2 (\hat{\psi} - \hat{y}) = 0 \quad [3]$$

where

$$\alpha^2 \equiv \frac{1 - e^{-(\hat{y} - \hat{\psi})^2}}{(\hat{y} - \hat{\psi})^2} \quad [4]$$

with $\hat{y} = y / R_o$ and $\hat{\psi} = \psi / R_o$. Here $\varsigma = k_\beta z$, $x = \omega_0 \tau$ and

$$\varepsilon \equiv \frac{k^2}{k_\beta^2 \omega_o \tau_o} \quad [5]$$

where $\tau_O$ is the neutralization time of the background gas and is approximately given by

$$\tau_o \cong 10^{-9} / P_{torr} \text{ sec} \quad [6]$$

$k^2$ is the coupling strength given by (I is the beam current and $I_O$ is ≈ 17 kA)

$$k^2 = \frac{2I}{\gamma \beta I_o R_o^2}, \quad [7]$$

and $\omega_O$ is the ion (angular) "sloshing" frequency in the field of the beam

$$\omega_o^2 = \frac{2qI}{McR_o^2}. \quad [8]$$

Here M is the ion mass, q the ion charge and c the speed of light.

## 2 LINEARIZED EQUATIONS

If both y and ψ are small compared to $R_o$ equations [2] and [3] may be linearized as

$$\frac{\partial^2 y}{\partial \varsigma^2} + y + \varepsilon x (y - \psi) = 0 \quad [9]$$

$$\frac{\partial^2 \psi}{\partial x^2} + \psi - y = 0. \quad [10]$$

We will solve these equations for a "tickler" excitation, that is

$$\frac{\partial y(0, x)}{\partial \varsigma} = .01 \sin(x) \quad [11]$$

$$y(0, x) = \psi(0, x) = \frac{\partial \psi(\varsigma, 0)}{\partial x} = 0. \quad [12]$$

These equations are appropriate for a beam and channel system that are characterized by a single betatron and ion "slosh" frequency. We are treating the case of solenoidal focusing that we assume is dominant compared to the focusing provided by the ion channel. Under this condition it is a good approximation to neglect the spread in betatron frequency that will result due to the non-linearity of the beam-channel force which arises from the non-uniform spatial profile of the channel.

However, it is not a good approximation to ignore the spread in ion resonance frequencies which arises from the non-uniform spatial profile of the beam. To account for this spread we use the "spread mass" model [3]. We modify the model by splitting the channel centroid into "filaments" labeled by a subscript λ which characterizes the frequency of a particular filament. Equation [10] is thus modified as

$$\frac{\partial^2 \psi_\lambda}{\partial x^2} + \lambda(\psi_\lambda - y) = 0. \quad [13]$$

The position of the channel centroid is then found by averaging the individual positions of the filaments over a distribution function

$$\overline{\psi} = \int f(\lambda)\psi_\lambda d\lambda. \quad [14]$$

For numerical work we will use the conventional definition and take a "top hat" distribution where

$$f(\lambda) = \frac{1}{\theta} \text{ for } 1-\theta \leq \lambda \leq 1. \quad [15]$$

For analytic work we will use a Lorentzian distribution (and equation [13] with $\lambda^2$ instead of λ):

$$f(\lambda) = \frac{\delta/\pi}{(\lambda-1)^2 + \delta^2} \quad [16]$$

where δ is the half-width of the distribution and the range of λ is from −∞ to +∞.

## 3 EFFECTS OF ENVELOPE OSCILLATIONS

We now wish to investigate the effects of an envelope mismatch in the accelerator on the growth of the instability. Since the dominant focusing for the beam is provided by the solenoidal field, an envelope mismatch will result in a beam radius that varies as

$$r_b = a(1 + \mu \sin 2\varsigma) \quad [17]$$

where we have assumed a particular choice of phase for the envelope oscillations without loss of generality. Because the channel is formed by the beam we can expect that there will be a similar variation for the channel radius. Thus the ion resonance frequency will be periodically varying in z. This is analogous to the case of "stagger tuning" the resonant frequency of cavities to detune the beam breakup instability.

We will investigate this effect by averaging over the fast ζ and x oscillations of both the channel and beam centroid positions [4].

First we write the factor g(ζ) as

$$g(\varsigma) = 1 + \mu \sin 2\varsigma. \quad [18]$$

Equations [9] and [13] then become

$$\frac{\partial^2 y}{\partial \varsigma^2} + y + \frac{\varepsilon x}{g(\zeta)}(y - \overline{\psi}) = 0 \quad [19]$$

and

$$\frac{\partial^2 \psi_\lambda}{\partial x^2} + \frac{\lambda^2}{g(\zeta)}(\psi_\lambda - y) = 0. \quad [20]$$

Then we may use the Laplace transform method (transforming in x to s and back again) along with equations [14] and [16] to obtain

$$\overline{\psi} = \int_0^x dx' \frac{y(\zeta,x')}{\sqrt{g(\zeta)}} e^{-\delta\frac{(x-x')}{\sqrt{g(\zeta)}}} \left[\sin\frac{(x-x')}{\sqrt{g(\zeta)}} + \delta\cos\frac{(x-x')}{\sqrt{g(\zeta)}}\right]$$

. [21]

We now write y(ζ,x) as

$$y(\zeta,x) = A(\zeta,x)e^{i(\zeta-x)} \quad [22]$$

where A is regarded as a slowly varying amplitude such that

$$\left|\frac{1}{A}\frac{\partial A}{\partial \varsigma}\right| << 1; \left|\frac{1}{A}\frac{\partial A}{\partial x}\right| << 1. \quad [23]$$

Treating μ as a small parameter, averaging equations [19] and [21] in ζ over 2π, we find after considerable algebra

$$i\frac{\partial A}{\partial \varsigma} + \frac{\varepsilon x}{2}\left[A - \frac{i}{2}\int_0^x dx' A(\zeta,x')J_o\left[\frac{\mu}{2}(x-x')\right]e^{-\delta(x-x')}\right] = 0$$

. [24]

By Laplace transforming equation [24] in x to s and using the method of steepest descents we find the asymptotic growth rate

$$y \propto e^{\Gamma(k_o,s_o)} \quad [25]$$

where

$$\Gamma(k_o,s_o) \approx -\delta x + x\left[-\frac{\mu^2}{8} + \sqrt{\frac{\mu^4}{64} + \left(\frac{\varepsilon\zeta}{2}\right)^2}\right]^{1/2}. \quad [26]$$

The exponential growth given by equations [25] and [26] is shown in Fig. 1.

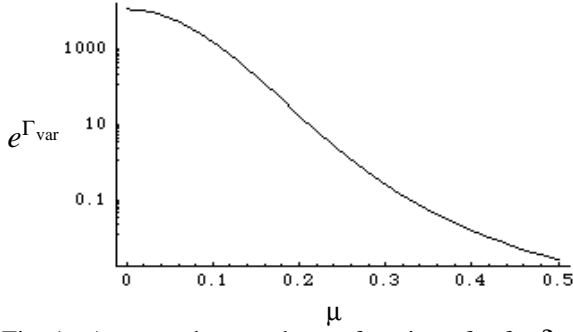

Fig. 1. Asymptotic growth as a function of µ for δ =.05.

We see that a small envelope mismatch can significantly reduce the linear growth, particularly for a large growth rate.

## 4 NON-LINEAR DEVELOPMENT

It is clear from equations [2] through [4] that when the beam and channel displacements become of order $R_o$ the beam-channel force falls off significantly as compared to the linear approximation used in equations [9] and [10].

We now extend equation [13] into the non-linear range (for a Lorentzian distribution) as

$$\frac{\partial^2 \hat{\psi}_\lambda}{\partial x^2} + \lambda^2 \alpha_\lambda^2 (\hat{\psi}_\lambda - \hat{y}) = 0. \quad [27]$$

By taking $\hat{y} = A(\zeta,x)e^{i(\zeta-x)}$ and $\hat{\psi}_\lambda = B_\lambda(\zeta,x)e^{i(\zeta-x)}$ with A and B both slowly varying we may average equations [2] and [27] to obtain (assuming B>>A)

$$2i\frac{\partial A}{\partial \zeta} + \varepsilon x \frac{(A-B)}{1+\frac{|B|^2}{4}} = 0 \quad [28]$$

$$-B_\lambda + \lambda^2 \frac{(B_\lambda - A)}{1+\frac{|B_\lambda - A|^2}{4}} \cong 0. \quad [29]$$

Equation [29] can be solved iteratively and integrated with equation [16] to find B. This result can be used to manipulate [28] into the form

$$\frac{\partial \psi}{\partial \xi} - \frac{2\psi}{1+2\delta\frac{\psi}{\psi_s}+\frac{\psi^2}{\psi_s^2}} \cong 0 \quad [30]$$

with $\quad |B|^2 \cong 8\delta\frac{\psi}{\psi_s} / \left(1+\frac{\psi^2}{\psi_s^2}\right) \quad [31]$

where $\psi \equiv |A|^2$, $\psi_s \equiv 32\delta^3$, and $\xi \equiv \varepsilon\zeta x/4\delta$, the number of e-folds of linear growth.

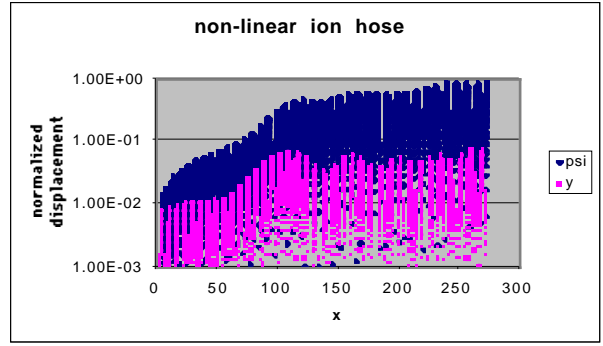

Fig. 2. Channel and beam position vs. x at the end of the accelerator from numerical solution. The blue curve is the channel centroid while the purple curve is the beam centroid. The top hat distribution (equation [15]) was used for θ=0.59.

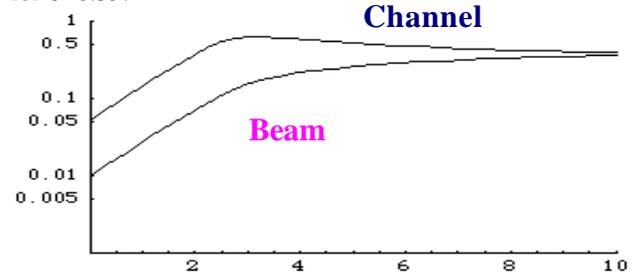

Fig. 3. Channel and beam position vs. ξ from equations [30] and [31] with δ=0.0939. This maximum value of the abscissa corresponds to 10 e-folds of linear growth of the beam centroid position, as is the case for figure 2.

## 5 CONCLUSIONS

We have shown that envelope oscillations that lead to a periodic detuning of the ion resonant frequency significantly reduce the linear growth rate of the instability. In addition, when the amplitude of the ion channel motion becomes of the order of $R_o$, the beam-channel force falls off significantly from the linear approximation. The betatron motion of the beam/channel causes a periodic modulation of the ion resonant frequency which increases the effective damping of the oscillations. This effect leads to the saturation of the beam centroid displacement at an amplitude that is of the order of 2δψ or about 10 - 30% of the channel amplitude.

## 6 ACKNOWLEDGMENTS

## 7 REFERENCES

[1] R. J. Briggs, private communication.
[2] H. L. Buchanan, Phys. Fluids **30**, 221 (1987).
[3] E. P. Lee, Phys. Fluids **21**, 1327 (1978).
[4] G. V. Stupakov, et. al., Phys. Rev. E **52**, 5499 (1995).